\documentstyle[aps,epsf]{revtex}
\global\firstfigfalse
\global\firsttabfalse
\begin{document}
\bibliographystyle {prsty}

\def\be{\begin{equation}}
\def\ee{\end{equation}}
\def\la{\langle}
\def\ra{\rangle}
\def\be{\begin{equation}}
\def\ee{\end{equation}}
\def\beq{\begin{eqnarray}}
\def\eeq{\end{eqnarray}}
\def\la{\langle}
\def\ra{\rangle}
\def\ul{\underline}
\def\td{thermodynamic}
\def\msc{mesoscopic}
\def\temp{temperature}
\def\gc{grandcanonical}
\def\parm{parameter}
\def\sm{statistical mechanics}
\def\bmtp{\bar\mu,T,\Phi}
\def\rarrow{\rightarrow}
\def\mio{\mu^{(i)}_0}
\def\mi{\mu^{(i)}}
\def\i{{(i)}}
\def\mno{\medskip\noindent}
\def\sno{\smallskip\noindent}
\def\bno{\bigskip\noindent}

\title{Statistical Ensembles and Spectral Correlations
 in Mesoscopic Systems}

\author{Alex Kamenev$^1$ and Yuval Gefen$^2$ }
\address{$^1$Institute for Theoretical Physics
University of California, Santa Barbara, CA 93106-4030, USA\\
$^2$Department of Condensed Matter Physics, The Weizmann Institute of Science,
Rehovot 76100, Israel. 
}

\maketitle

\baselineskip 18pt\noindent

\begin{abstract}
Employing different statistical ensembles may lead to qualitatively different
results concerning averages of physical observables on the mesoscopic scale.
Here we
discuss differences between the canonical and the grandcanonical ensembles due
to
both quenched disorder {\it and} thermodynamical effects. We show how these
differences are related to spectral correlations of the system at hand, and
evaluate
the conditions (temperature, system's size) when the thermodynamic limit is
achieved. We demonstrate our approach by evaluating the heat capacity,
persistent
currents and the occupation probability of single electron states, employing a
systematic diagrammatic approach.
\end{abstract}\par

\mno
\section{Introduction}

Standard textbooks on equilibrium statistical mechanics often assert that the
various statistical ensembles in standard use (e.g., the canonical ensemble
(CE) and
the grandcanonical ensemble (GCE) are equivalent. Landau and Lifshits [1]
assert
hat ``$\dots$ all three distributions, the microcanonical and the two forms of
the
Gibbs distribution, are in principle suitable for determining the
thermodynamic
properties of the body. The only differnce from this point of view lies in the
degree of mathematical convenience". Huang [2] claims that ``$\ldots$ the
grand
canonical ensemble is trivially equivalent to the canonical ensemble for N
particles", and there are many more examples. These assertions should be
understood
as referring to average thermodynamic quantities (as distinct from
fluctuations; it
is evident that by the very nature of the respective ensembles, fluctuations
in the
number of particles, for example, are very different between the CE nd the
GCE).
Moreover, it is assumed that the body under consideration is macroscopic
(i.e., the
``thermodynamical limit" has been taken). In this limit the relative
fluctuations
tend to zero.
\par It is important to stress, though, that thermodynamic (or
statistical-mechanical) quantities may be perfectly defined even for a system
with
few degrees of freedom -- provided they are in equilibrium with an infinite
reservoir (heat bath, particle reservoir, etc.). A well-known example is the
Fermi-Dirac distribution function which is obtained by considering a
single-level
system coupled (weakly) to an electron bath at temperature T and chemical
potential
$\mu$. In the limit of small systems fluctuations are large (relative to mean
values). As we have stressed above, their nature is very different between the
different ensembles. More interestingly, average thermodynamic quantities too
may be
different between the CE and the GCE. Most strikingly, qualitative differences
between the two ensembles may arise when quantum coherence comes into play, on
mesoscopic scales and downwards. Novel effects may appear, which are quite
remote
from our common wisdom and everyday intuition as applied to macroscopic
systems in
the thermodynamic limit. The thermodynamic limit is attained as the system's
size
tends to infinity and/or when the temperature is made sufficiently high.
\par It is evidently interesting to identify the scale of temperature (in
conjunction with the system's size) at which the non-trivial differences
(i.e.,
differences in mean values, possibly qualitative) between the CE and the GCE
are
suppressed. Furthermore, we shall note that on finite length or temperature
scales
there are certain differences between thermodynamic relations and
statistical-mechanical ones (therefore these two different frameworks may lead
to
different predictions, although, as was noted above, both statistical
mechanics and
thermodynamics are well-defined even in the context of finite systems.
Consequently
there are characteristic temperature or length scales which mark the
suppression of
the differences between statistical mechanics and thermodynamics, and the
crossover
to the ``thermodynamic limit".
\par It should be stressed that in most practical instances the fabrication of
samples is not a perfectly reproducible process. Commonly there are
sample-to-sample
variations in sample size and shape, in the mesoscopic impurity configuration
and
the location defects (we shall refer to all these factors as ``disorder").
Evidently
it is possible to ask questions pertaining to sample specific observables, and
to
evaluate the magnitude of sample-to-sample fluctuations in these observables.
But
it is also of interest to consider {\it ensemble averaged} quantities, where
averaging over disorder is included (the concept of ensemble will be
elucidated
below). Such ensemble averaged quantities will be at the focus of the present
study.
\par The purpose of this paper is twofold. We shall first review and expand on
the
differences between the CE and the GCE (in the context of ensemble averaged
quantities) both due to quenched disorder and to thermodynamic fluctuations.
The
latter is manifest only at finite temperatures. This part includes careful
examination of the notion of these ensembles in conjunction with the proper
averaging procedures. Secondly, we shall address certain questions central to
our
understanding of statistical mechanics and \td s on the mesoscopic scale. Thes
questions include the following points: (1) We shall identify the (length
scale
dependent) \temp\ over which differences between the two ensembles are
suppressed.
(2) We shall relate this to the \temp\ (or length) scale over which
differences
between statistical mechanics and \td s are suppressed. We shall thus identify
the scales at which the \td\ limit is approached. (3) We shall relate the
above to
energy scales characterizing correlations of the underlying energy spectrum.
These
concepts will be illustrated below.
\par The scope of the questions addressed here is quite general and includes
generic
many body systems. Our {\it quantitative} discussion, though, will be carried
out for
systems of noninteracting electrons.
\par The outline of this paper is as follows.
\par In Section 2 we discuss various definitions of ``canonical" and
``grandcanonical". We distinguish between fluctuations in the number of
particles
(electrons), $\delta N$, which are due to \td al effects, as opposed to
quenched
disorder induced $\delta N$. We also discuss the difference between canonical
(or
grandcanonical) constraints, defined with respect to the way the system is
prepared
(at equilibrium), and such constrains as defined when the system under
consideration
is subject to a dynamical perturbation. We next explain how to express
ensemble
averaged canonical quantities in terms of ensemble averaged \gc\ quantities
and
fluctuations thereof. This is done at zero temperature and is presented as an
expansion to an arbitrary order in a small \parm.
\par Section 3 contains a discussion of the differences between the CE and the
GCE
at finite \temp. This is related to differences between \td s and \sm\ on the
\msc\
scale. Exact \td\ identities should be corrected when finite systems (at
finite \temp
s) are considered {\it within the framework of \sm}. This further modifies the
relations between the CE and the GCE found above at T=0 alluded to above. We
find how
these differences between \td\ identities and statistical mechanical relations
depend on
\temp\ (and system's size) and evaluate the scale over which crossover to the
``\td\
limit" takes place. It turns out that this scale (say, in \temp) is usually
much
larger than what one might normally guess (namely, that it is of the order of
the
level spacing). We demonstrate our approach by evaluating the manner in which
three
different physical observables depend on the type of the ensemble employed;
these
observables are the heat capacity, the persistent current and the single level
occupation
probability. Differences between \gc ly and canonically averaged quantities
(e.g., the
Fermi-Dirac distribution function and its canonical counterpart) are due to
both
quenched disorder and \td\ fluctuations.
\section{Statistical Ensembles and Quenched Disorder: T=0}\par
\mno{\bf 2a~~~What Do We Mean by ``Canonical" and ``Grandcanonical"
Conditions:
Particle Fluctuations and Various Averaging Procedures}

The notion of {\it ensemble averaging} is in the very heart of \sm. Rather
than
considering a particular time evolution of a specific system with given
initial
conditions, we consider an ensemble of systems (or a set of initial
conditions) and
average over them with appropriate weights, stipulating that in any given set
of
experiments (or in an experiment that lasts a finite period of time) the
experimental
conditions are generally not reproducible to the last minute detail, but
rather
random sampling of the systems which form the ensemble takes place.
\par The concept of a \gc\ ensemble (as opposed to a canonical ensemble)
invokes the
notion of fluctuations in the number of particles, $\delta N$. (In our case we
shall
refer to fluctuations in the numbe of electrons in the system). We now
describe four
different thought experiments which help us to eliminate some confusion that
arises
in the literature regarding what really is meant by CE or GCE.\newline
\sno {\bf Experiment 1:} We consider a system (in our context a \msc\
conductor with
disorder) coupled {\it weakly} to an external reservoir at \temp\ T and a
chemical
potential $\mu$ (Fig. 1a). The system will reach equilibrium with the external
reservoir (if the coupling is weak the equilibration time may be long, but we
are
not interested here in transients). Even when equilibrium is attained the
number of
electrons in the system may still fluctuate in time; at finite \temp\ the
system may
exchange electrons with the reservoir. Denoting the number of electrons at
times [3]
$t_1,t_2,\ldots t_n$ by $N_1, N_2,\ldots N_n$ we obtain a fluctuating sequence
of
numbers, whose variance is $\delta N^{(1)}$.\newline
\sno {\bf Experiment 2:} We consider {\it identical microscopic replica} of
the same
\msc\ systems, all weakly coupled to the same external reservoir (Fig. 1b).
Measuring the number of electrons in each system (at equilibrium) at a given
arbitrary time we obtain a fluctuating sequence $N_1,N_2,\ldots N_n$ whose
variance
is denoted as $\delta N^{(2)}$.
\par By the {\it ergodic hypothesis} $\delta N^{(1)}=\delta N^{(2)}$.
Moreover, they
both tend to zero at $T\rarrow 0$. This reflects the fact that the origin of
these
fluctuations is {\it dynamical}: the Hamiltonian of the system contains a
(small)
coupling term to the reservoir. The number of electrons in the system is,
therefore,
not a conserved quantity. Evidently in the limit of $T\rarrow 0$, when the
level
broadening in the system due to the coupling term is smaller than the level
spacing,
and if $\mu$ is chosen to fall within the ``gap" between two consecutive
levels, the
exchange of electrons with the reservoir is suppressed.\newline
\sno {\bf Experiment 3:} We now consider an ensemble of \msc\ conductors which
are
{\it macroscopically} similar (they are made of the same material, are
characterized
by the same volume and impurity concentration, etc.) but are {\it
microscopically}
distinct (small variations in shape; different impurity configurations (Fig.
1c)).
All these systems are weakly coupled, and equilibrated, with the same external
reservoir. As a result, to each member of this ensemble corresponds a different
spectrum (cf. Fig. 2a). Thus, even at T=0 (and with every member possessing
the same
value of chemical potential) different members of the ensemble will have
different
numbers of electrons. This defines $\delta N^{(3)}$ which does not vanish even
in
the $T\rarrow 0$ limit. The origin of this $\delta N^{(3)}$ is {\it quenched}
disorder.\newline
\sno {\bf Experiment 4:} We may consider a single member of the ensemble at
equilibrium, as function of an externally controlled parameter (magnetic
field,
Aharonov-Bohm (AB) flux, electric field, external pressure, etc.) (Fig. 1d).
We
shall denote this external \parm\ by $\Phi$. Generally the spectrum of the
system
depends on $\Phi$. A generic situation is depicted in Fig. 2b. The number of
particles (occupied levels) at equilibrium varies with $\Phi$ (e.g., at T=0
$N(\Phi_1=2$, $N(\Phi_2)=3$, cf. Fig. 2b), which defines $\delta N^{(4)}$.
Again,
this quantity does not vanish even in the limit $T\rarrow 0$. The order of
magnitude
of $\delta N^{(3)}$ and $\delta N^{(4)}$ may be very different. For diffusive
systems
with $\Phi$ representing an AB flux it has been found by Altshuler
and Shklovskii [4,5] that for a two-dimensional systm (d=2) and hard core
scatterers
$\delta N^{(3)} \sim \hbar/E_c\tau$ (for $(\frac{L}{\ell})^2 \gg 1)$
where $\tau$ is the elastic mean free time
and $E_c={\hbar D\over L^2}$ is the Thouless energy ($L$ is the system's
linear
size; $i\ell$ is the elastic mean file path,
$D$ is the diffusivity). By contrast [6] $\delta N^{(4)}\sim 1$.
\par In the first part of this paper we shall set the \temp\ to be zero. We
shall
thus suppress any dynamic fluctuations in $\delta N$, and the distinction
between
the GCE and the CE will be based on whether or not there are fluctuations in
the
number of particles due to quenched disorder. (Only at a later stage shall we
also
consider finite \temp s). Thus, in the present context we shall employ the
term CE
to refer to a set of finite systems whose (sample specific) number of
electrons
$N^i$ ($i$ is used here as an index for the ensemble member) is selected
independently of the disorder dependent energy spectrum, and is specifically
$\Phi$ independent. Similarly, a GCE implies that the sample specific
chemical potential, $\mu^i$, is selected independently of the disorder energy
spectrum, and is specifically $\Phi$ independent. Throughout most of this
paper we
shall consider systems of noninteracting electrons, although some of the
notions
here may be generalized to include systems of interacting particles.
\par It is possible to refer to a {\it strong} CE where $N^i=N$ for all $i$,
and
similarly to a {\it strong} GCE where $\mu^i=\mu$ for all $i$ (these were
implicitly
the conditions in Refs. 6-10 and Ref. 4, respectively). Alternatively,
averaging
within the CE (GCE) may be realized by randomly selecting values for
$N^i(\mu^i)$
out of a broad and smooth distributions. We shall refer to these as {\it weak}
CE
(GCE) respectively. Below we shall argue that in many generic situations
strong and
weak averaging scenarios, as defined above, are equivalent, and will exploit
this
equivalence to our advantage. Figs. 3a,b depict schematically typical strong
CE and
strong GCE scenarios respectively. In Fig. 3a the probability that $\mu$ falls
in a
given interval between two levels is proportional to the distance (``gap")
between
these two levels. The effective (sample specific) chemical potential in the CE
($T\gg\Delta$) is located half way between the last occupied ($N^{\ul{th}}$)
level
and the first vacant ($N+1^{\ul{st}}$) level [11], Fig. 3b, regardless of the
size
of the interlevel gap. Thus, compared with the GCE, there is an enhanced
probability
to find  narrow gaps in the vicinity of the effective chemical potential in
the CE.
The implications of this observation will become apparent below.
\par The distinction among various types of ensembles may be carried further
if we
note that the nature of the ensemble depends conceptually on both the way it
is
prepared (brought into equilibrium), and the conditions under which the
system is
allowed to respond to an external bias. To be specific, let us consider an
ensemble
of multiply connected Aharonov-Bohm rings, subjected to magnetic flux, $\Phi$,
threading each of them. At the preparation stage (which takes place at flux
$\Phi$),
the rings may be brought into equilibrium either subject to canonical
conditions
($\{N^i\}$ are chosen to be random independent \parm s, i.e., the number of
electrons at
each ring is selected independently of the ring's specific impurity
configuration),
or in accordance with grandcanonical conditions $\{\mu^i\}$ are chosen
independently
of the impurity configuration). Subsequently, one performs measurements (e.g.
of the
ring's magnetic moments) at the flux $\Phi$. Then, too, as $\Phi$ is varied
(to
obtain the response of the system), either canonical (number of electrons is
kept
unchanged) or \gc\ (value of chemical potential is set fixed) conditions may
be
adopted. According to this classification, where a distinction is made between
the
preparation and the measurement stages, the CE is now denoted as
canonical-canonical
(C-C) and the GCE as GC-GC. We may also consider an intermediate, ``hybrid"
type of
ensemble, GC-C. This ``hybrid" procedure implies that the number of particles
remains unchanged during the time $\Phi$ is varied. This situation is achieved
either when the system is decoupled from the reservoir before $\Phi$ is
varied, or if the
coupling strength,
$\gamma$, (affecting the inelastic broadening of the levels) [12] satisfies
$\gamma<\omega_{bias}$ where $\omega^{-1}_{bias}$ is the characteristic time
scale
for time modulations of the external bias. For $\gamma>\omega_{bias}$ the
response
to the external bias may be referred to as \gc. We note that as long as we
allow for
(weak) coupling to the particle bath (which is necessary at the stage the
system is prepared \gc ly) and do not turn it off completely, the GC-C
procedure is
not an adiabetic one: when the rate of change of $\Phi$ is slower than any
other
time scale in the system the procedure converges toward a GC-GC procedure.
\par Following the above discussion and referring to the nature of both the
preparation and measurement stages, we may consider GC-GC, C-C or GC-C
procedures.
At least in principle, one may realize these various ensembles experimentally
[13,14]. The fourth combination, C-GC, is not physically relevant. Evidently
the
procedure we take at any given stage needs not be purely canonical or \gc. For
example,we
may select the number of electrons within each ring to be only slightly
correlated
with the ring's disorder configuration. Thus there is a whole continuum of
statistical procedures.\newline
\mno{\bf  2b~~~Relations Among Various Statistical Ensembles: T=0}
\par Returning now to the two principle ensembles (CE and GCE, or, employing
our
alternative notation, C-C and GC-GC), it is desirable  to find a relation
between
averages taken within the two ensembles respectively. From the calculational
point of
view, the GCE is more amenable to analytical studies (employing, e.g.,
disorder Green
function techniques, random matrix theory (RMT), or supersymmetric (SUSY)
calculations).
\par To derive a general relation between GCE and CE averages we consider a
strong GCE
situation. The (thermodynamic average) number of electrons in the $i^{th}$
member of
the ensemble is given by
\be N^i(\Phi,\mu)=\int\!d\epsilon\nu^i(\epsilon,\Phi) f(\epsilon-\mu)\;,\ee
where $\mu=const.$\ and $\nu^i(\epsilon,\Phi)$ is a sample and flux dependent
density of
states (to be more concrete we presently assign to $\Phi$ the meaning of an
external
magnetic flux). We define the ensemble averaged number of particles
\be \bar{N}(\mu)\equiv\la N^i(\Phi,\mu)\ra\;.\ee
The flux dependence of $\bar N(\mu)$ is exponentially weak, hence we ignore
it. We also
define the mean level spacing as
\be \Delta^{-1}\equiv{\partial\bar N\over\partial\mu}\;.\ee
We further employ the approximation
\be {\partial^k\bar N\over\partial\mu^k}\equiv 0 \qquad \hbox {for}\; k\ge
2\ee
This is exact in two dimensions ($d=2$) and involves corrections small in
$\Delta/\mu$
in $d\ne 2$.
\par For strong CE one has
\be N=\int\!d\epsilon\nu^i(\epsilon,\Phi) f(\epsilon-\mu^i(\Phi))\;,\ee
where $N\equiv const.$\ and $\mu^i(\Phi)$ is a sample and flux dependent
chemical
potential. Next we define the ensemble averaged chemical potential, $\bar\mu$,
as the
solution of the following equation:
\be N=\bar N(\bar\mu)\;.\ee
We then write the chemical potential as
\be \mu^i(\Phi)=\bar\mu+\delta\mu^i(\Phi)\;.\ee
Expanding now the Fermi function in Eq. (5) and using the GCE relation, Eq.
(1), one
obtains
\be N=\sum^\infty_{k=0}{1\over
k!}\left[{\partial^k\over\partial\bar\mu^k}N(\Phi,\bar\mu)\right]
(\delta\mu^i(\Phi))^k\;.\ee
Employing Eqs. (3), (4) and (6), we obtain the following transcendental
equation for
the dimensionless variable, $X\equiv\delta\mu^i(\Phi)/\Delta$:
\be X+\sum^\infty_{k=0}\Delta^kX^kC_k=0\;,\ee
where
\be C_k\equiv {1\over k!}{\partial^k\over\partial\bar\mu^k}\delta N\,;\qquad
\delta
N\equiv N^i(\Phi,\bar\mu)-\bar N(\bar\mu)\, .\ee
\par We shall look for the solution of Eq. (9) in the form of the series
\be X=\sum^\infty_{n=0}a_n\Delta^n\;.\ee
Substituting Eq. (10) into Eq. (9) and comparing respective orders of powers
of
$\Delta$ we find
\be a_n={(-1)^{n-1}\over n!}\left(\delta N^n{\partial\delta
N\over\partial\bar\mu}\right)^{(n-1)}\;,\ee
where $(n-1)$ denotes the $(n-1)^{st}$ derivative with respect to $\bar\mu$.
Finally
one has a solution for $\delta\mu^i(\Phi)$
\be \delta\mu^i(\Phi)\equiv\Delta\sum^\infty_{n=0}{(-1)^{n-1}\over n!}\Delta^n
\left(\delta N^n{\partial\delta N\over\partial\bar\mu}\right)^{(n-1)}\;
.\ee
This is a central result of our analysis so far, which enables us to calculate
various
canonical quantities. As an example we consider the situation where $\Phi$ is
an AB
flux and work out the canonically averaged persistent current
\be I^i_{CE}(\Phi) =\sum^\infty_{n=0} {1\over n!}{\partial^n
I^i_{GCE}\over\partial\bar\mu^n}(\delta\mu^i)^n\, .\ee
Employing a standard \td\ relation
\be{\partial I_{GCE}\over\partial\mu}\equiv{\partial
N^i(\Phi,\bar\mu)\over\partial\bar\Phi}\;,\ee
and substituting Eq. (13)  into Eq. (14) we finally obtain
\be
I_{CE}(\Phi)=I_{GCE}(\Phi)-\Delta{\partial\over\partial\Phi}
\left[\sum^\infty_{n=0}{(-1)^n\over (n+2)!}\Delta^n
{\partial^n\over\partial\bar\mu^n}\big[\delta N^{n+2}\big]\right]\;.\ee
One immediately notices that all terms in this series, except the $n=0$, have
the form
\be \Delta{\partial\over\partial\bar\mu}(\cdots)\; .\ee
This means that once averaged they yield a results proportional to
$\Delta/\bar\mu$, as
the only dependence of the averaged quantities for $\bar\mu$ arises from the
diffusion
constant $D=\bar\mu\tau/m$ (we are in $d=2$, so the average density of states
is
constant).
\par Equation (16) is a prototype of a relation connecting CE to a GCE average. It
goes
beyond the results of refs. 6-10 in that it provides a systematic way to
obtain higher
corrections. More generally Eq. (16) can be cast in the form
\be {\partial
F\over\partial\Phi}\bigg)_{CE}={\partial\Omega\over\partial\Phi}\bigg)_
{GCE}+\Delta{\partial\over\partial\Phi} \sum^\infty_{n=0} {(-1)^n\over(n+2)!}
\Delta^n
{\partial^n\over\partial\bar\mu^n}\big[\delta N^{n+2}\big]\ee
where $\Phi$ now represents any static externally controlled parameter. Note
that here
$\delta N$ is a \gc\ quantity, so the canonical quantity  on the l.h.s. of Eq.
(18) is
expressed solely by means of GCE quantities.
\par A simple version of Eq. (18), namely this relation where only the $n=0$
term on
the r.h.s. has been included, has been extensively employed [6-10] to study
differences
between the CE and the GCE persistent current in normal (as opposed to
superconducting)
rings and cylinders [15], and the anomalous paramagnetic orbital magnetism of
small
dots [16]. It is also interesting to consider the problem of ``hybrid"
ensembles,
mentioned above [13,14,17].
\section{Statistical Ensembles: $T>0$}
So far we were interested in differences between the ``canonical" and the
``\gc"
ensembles (and hybrids thereof) due to quenched disorder statistical
fluctuations.
These differences are clearly manifested at T=0. The CE was represented as an
ensemble
of effectively \gc\ systems, with sample specific (and flux dependent)
chemical
potentials. It is argued below that this is not the case at $T>0$: even on the
level of
a specific system, and even when we consider thermodynamically averaged
quantities (and
not fluctuations), a canonical system cannot be replaced by an effective \gc\
system.
This will be the source to further differences between the CE and the GCE
(i.e.,
differences beyond those discussed in Refs. 6-10,18). Here we study these
difference
and show that they are related to differences between \sm\ and thermodynamics
on the
mesoscopic scale.
\par In other words, consider an observable $A$. Canonical and \gc\ averages
will be
denoted as $\la A^i|_N\ra$ and $\la A^i|_\mu\ra$ respectively. The difference
between
such averages can be written as
\be \la A^i|_N\ra-\la A^i|_\mu\ra =\la A^i|_{\mu^i(N)}-A^{\i}|_\mu\ra +\la
A^i|_N-A^{\i}|_{\mu^i(N)}\ra \;.\ee
The first term was the center of our discussion in Section 2. Note that if one
assumes
the relation
\be{\partial F^{(i)}\over\partial x}\big|_N
={\partial\Omega^{(i)}\over\partial
x}\left|_{\mu(N,\Phi)}\right.\ee
(where $\Phi$ is an externally controlled parameter) this implies that the
second term on
the r.h.s. of Eq. (19) vanishes. The first term on the r.h.s. exhausts, in
that case,
the difference between the CE and the GCE, including the temperature
dependence. In
general, though, as is discussed below, Eq. (20) is not satisfied exactly at
$T>0$,
implying the non-vanishing of the second term on the r.h.s. of Eq. (19). The
reason is
that at finite temperature the relation given in Eq. (20) above, while being
exact
within the framework of thermodynamics, includes correction terms when
considered within
the framework of \sm\ and applied to finite systems (see Section 3b). It
follows that
the fact that at finite temperatures we allow for dynamical particle
fluctuations in
the \gc\ case (while no particle exchange with the reservoir is allowed under
canonical
constraints) affects quantities related to fluctuation terms, e.g. $\la\delta
N^2\ra$.
In particular this applies to the relations between GCE and CE averages, which
contain
fluctuation terms (cf. Eq. (18)). It is important to stress that the finite
\temp\
difference between canonical and \gc\ quantities, which is of a dynamical
nature, exists
even for sample specific quantities. This contribution is to be superimposed
on top of the contribution discussed earlier, which arises due to averaging
over quenched
disorder. \par
\mno
{\bf 3a~~~Differences Between the Canonical and the Grandcanonical Ensembles:
Single
Particle Level Occupancy}
\par We demonstrate the $T>0$ effect by studying the difference in the single
particle
level occupancy between canonical and \gc\ systems. As a simple example, let
us consider
first a two-level system occupied by $N=1$ electron (in the \gc\ case this
becomes
$\bar N=1$). This is depicted in Fig. 4. Here $\Delta$ denotes the level
spacing. At T=0
there is no practical  difference between the canonical and the \gc\ system.
By
contrast, at $T>0$, the probability of finding the second level occupied is
given by
\beq f_C\hbox{(level no. 2)}&=& {1\over e^{\beta\Delta}+1}\quad
canonical\nonumber
\\
f_{GC}\hbox{(level no. 2)}&=& {1\over e^{\beta\Delta/2}+1} \quad
grandcanonical\eeq
(in the canonical case we use the Boltzman factor for any given manybody
state,
projecting then onto the single particle state at hand; in the \gc\ case we
employ the
Fermi-Dirac factor, $f\equiv f_{GC}$. Clearly $f_C\ne f_{GC}$.
\par We next consider a many level system. Closely related works [19] which
include
analytical studies of uniformly spaced spectra [20], possibly with the effect
of level
degeneracy [21], and a numerical study [22], have been published earlier. The
Hamiltonian  of the system, $\hat H$, is assumed to preserve the number of
particles
(electrons), i.e. $[\hat H,\hat N]=0$. The canonical partition function is
given by
\be Z^{(i)}_C(N) = 
\mbox{Tr}\big(e^{-\beta \hat H^{(i)}}\delta (\hat N-N)\big)\ee
where the hatted quantities are operators, and the $Tr$ is unrestricted
(i.e., it
includes summation over eigenstates of $\hat N$ with various eigenvalues). We
use the
following representation of the Kroenecker delta function
\be \delta(\hat
N-N)=\beta\int\limits^{i\pi\over\beta}_{-i\pi\over\beta}\!{d\mu\over
2\pi i}   e^{\beta\mu(\hat N-N)}\;.\ee
Substituting Eq. (23) into Eq. (22) we obtain
\be Z_C(N)=\beta\int\limits^{i\pi\over\beta}_{-i\pi\over\beta}\! {d\mu\over
2\pi i}\,
e^{-\beta\mu N} \mbox{Tr}\, e^{-\beta(\hat H-\mu\hat N)}.\ee
Our goal now is to expand the r.h.s. of Eq. (24) around a saddle point
[23-25]. To
simplify the discussion we shall restrict ourselves here to systems of
noninteracting
fermions. We can find a diagonalizing basis in which $\hat H$ can be written
as
\be\hat H=\sum_k \epsilon^{(i)}_ka^+_k a_k\;. \ee
The canonical partition function is then written as
\beq Z^{(i)}_C(N) &=&
\beta\int\limits^{-i\pi\over\beta}_{-i\pi\over\beta}\!{d\mu\over
2\pi i}\, e^{-\beta\mu N}\prod_k \big(1+e^{-\beta(\epsilon^{(i)}_k-\mu}\big)
\nonumber
\\
&=&\int\limits^{i\pi\over \beta}_{-i\pi\over\beta}\!{d\mu\over 2\pi i}
e^{Ng(\mu)}\;
,\eeq 
where
\be g(\mu)=-\beta\mu +{1\over N}\sum_k \ln
n\big(1+e^{-\beta(\epsilon^{(i)}_k-\mu)}\big)\, .\ee
In analogy with the \gc\ ensemble, the single particle level occupation factor
in the
canonical case, $f^{(i)}_C(\epsilon_n;N)$ is given by
\be f^{(i)}_C(\epsilon^{(i)}_n;N)
=-{1\over\beta}{\partial\over\partial\epsilon^{(i)}_n} Z^{(i)}_C (N)=
{\int^{i\pi\over\beta}_{-i\pi\over\beta}\! d\mu  \,
f(\epsilon^{(i)}_n-\mu)e^{Ng^{(i)}(\mu)}\over
\int^{i\pi\over\beta}_{-i\pi\over\beta}\! d\mu\cdot
e^{Ng^{(i)}(\mu)}}\,.\ee
At this stage we shall replace the limits of the integrals by infinite
contours,
$\int^{i\pi\over\beta}_{-i\pi\over\beta}\rarrow \int^{+\infty}_{-\infty}$,
and evaluate these integrals employing the stationary phase approximation. It
is
possible to show [27] that this procedure is sound, and that the main
contribution to
the integral is contained within the original interval
$[{-i\pi\over\beta},{i\pi\over\beta}]$ when the temperature is larger than the
mean
level spacing.
\par
One may now expand the numerator and the denominator of the r.h.s.\ of
Eq.~(28) (or Eq.~(24)), making use of the diagrammatic expansion
outlined further below. Here we shall derive a simple approximate
expression, retaining only the Gaussian term in the exponent in
Eq.~(28), and expanding $f$ up to the second order $m \mu-\mu_0^{(i)}$
($\mu_0^{(i)}$ is the saddle point of the expansion, see Section~3b).
One then obtains (on the limit of a quasi continuous spectrum):
\be f_C(\epsilon;N)=f(\epsilon-\mu^{(i)}_0)-{\Delta\over 2\beta}\,
f''(\epsilon
-\mu^{(i)}_0)\;.\ee 
We readily see that the canonical distribution $f_C$ is qualitatively
equivalent to the
\gc\ distribution at a lower temperature. To find this effective lower
temperature we
write
\be
f'(\mu;\beta_{GC})=f'_C(\epsilon=\mu;N;\beta_C)=f'(\mu;\beta_C)-{\Delta\over
2\beta} f''(\mu;\beta_C)\, .\ee
This yields
\be \beta_{GC}\approx\beta_C\bigg(1+{\Delta\beta_C\over 4}\bigg)\;,\ee
where $\beta_{GC}$ is the effective \gc\ inverse temperature which corresponds
to a
canonical inverse temperature $\beta$. Based on the above discussion, the last
expression is valid for $T\simeq\Delta$. It can be rewritten as
\be T_{GC}\approx {T_C\over 1+{\Delta\over 4T_C}}\ee
which in the limit of $T_C\gg\Delta$ becomes
\be T_{GC}\approx T_C-{\Delta\over 4}\qquad T_C\gg\Delta\, .\ee 
One may employ slightly different criteria for defining the effective
temperature of
the canonical distribution function (e.g.,
$f(\mu,\beta_{GC})=f_C(\epsilon=\mu;N;\beta_C)$ instead of Eq. (30)). But the
statement
that the canonical distribution, at temperatures larger than $\Delta$,
corresponds to
$T_C>T_{GC}$ should remain valid. A similar shift of the effective temperature
has been
discussed (in the context of a uniform spectrum) in Refs. [20-22].
This also agrees qualitatively with the 2-level scenario discussed
above, where the effective canonical temperature is higher by a factor
of 2.
\par\mno
{\bf 3b~~~Differences Between the Canonical and the Grandcanonical Ensembles
at
$T>0$: A General Approach}
\par Our next step is to develop a systematic method for calculating physical
observables within the canonical ensemble, accounting for finite temperature
effects,
in particular for the fact that at finite temperature $f_C\ne f$. We
incorporate into
Eq. (24) for the canonical partition functin, $Z_C$, the relation
\be e^{-\beta\Omega(\mu,T,x)}=
\mbox{Tr}\,  e^{-\beta(\hat H-\mu\hat N),}\ee 
writing
\be Z_C(N,T,\Phi)\approx {\beta\over 2\pi i}
\int\limits^{+i\infty}_{-i\infty}\! d\mu
e^{-\beta(\Omega(\mu,T,\Phi)+\mu N)}\, ,\ee
The limits of the integral have been extended to infinity, following the same
arguments
as in the derivation of $f_C$. We note that the expression for $Z_C$ is quite
general and
is valid for systems of interacting electrons as well. We shall restrict
ourselves to
systems of non-interacting electrons in the diffusive regime. Our goal here is
to
evaluate ensemble averaged thermodynamic observables within the canonical
ensemble. (We
stree that similar ideas may be employed for calculating other classes of
observables,
e.g., linear response transport coefficients).
\par As we have done above, we define the sample specific saddle point of the
integral
in Eq.~(35), $\mu^{(i)}_0(T,\Phi)$ [28]
\be {\partial\over\partial\mu}\big[\Omega^{(i)}(\mu,T,\Phi)+\mu
N\big]\bigg|_{\mu^{(i)}_0(T,\Phi)}=0\, .\ee
Let us {\it define} the \gc\ expectation value of the particle number as
\be N^{(i)}(\mu,T,\Phi)=-{\partial\Omega^{(i)}(\mu,T,\Phi)\over\partial\mu}\,
.\ee
(For discussion of the relation between this identity, defined within
 statistical
mechanics, and \td\ identities, we refer the reader to Section 3d). The saddle
point
equation then assumes the form:
\be N^{(i)}\big(\mu^{(i)}_0(T,\Phi),T,\Phi\big)=N\, ,\ee
where $N$ is a given integer. We are supposed to solve Eq. (38) for
$\mu^{(i)}_0$, and
substitute it in the expression for $Z_C$, Eq. (35), expanding around the
saddle point:
\be \Omega^{(i)}(\mu)+\mu N= F^{(i)}\big({\mio}\big)+{1\over
2!}{\partial^2\Omega^{(i)}\over\partial\mu^2}\bigg|_{{\mio}}(\mu-{\mio})^2+\ldots
+{1\over n!}{\partial^n\Omega^{(i)}\over\partial\mu^n}(\mu-{\mio})^n +\ldots\;
. \ee
We note that ${\mio}$ is real. We have used the notation
$F^{{\i}}({\mio})\equiv\Omega^{{\i}}(\mu_0)+{\mio}N$. Unfortunately the above
expression,
Eq. (39) is not very useful for practical purposes. Each term is to be
evaluated at the
chemical potential ${\mio}$ which is a function of the external parameter
$\Phi$, but
even more significantly, is sample specific. Evidently, it is important to
perform
ensemble averaging by calculating the sample specific function ${\mio}(\Phi)$
for each
and every member of the ensemble. Instead we replace this quantity by a
(positive)
constant $\bar\mu$, assumed to be the result  of averaging ${\mio}(\Phi)$ over
$\Phi$
and over ensemble realizations,
$\bar\mu\equiv\la\overline{{\mio}(\Phi)}^\Phi\ra$. We
assume that in some sense (to be discussed below) $\bar\mu$ is close to
${\mio}(\Phi)$.
We thus distort the integration contour, $C$, of Eq. (35) which goes through
the saddle
point ${\mio}(\Phi)$ into another contour, $\bar C$, which goes through
$\bar\mu$.
Expanding around the latter, Eq. (39) is now replaced by
\be\Omega^{{\i}}(\mu)+\mu N=F^{{\i}}(\bar\mu)-{1\over 2\Delta} (\mu-\bar\mu)^2
-\sum^\infty_{n=1}\left[{\partial^{n-1}\over\partial\bar\mu^{n-1}}\delta
N^{{\i}}\right] (\mu-\bar\mu)^n\; .\ee 
Here $\delta N^{{\i}}\equiv N^{{\i}}(\bar\mu)-N$. The mean level spacing
$\Delta$
satisfies $\Delta=\left\la{\partial
N\over\partial\mu}\big|_{\bar\mu}\right\ra^{-1}$.
In writing Eq. (40) we assumed that $N={\bar\mu\over\Delta}$, such that
${\partial^n\over\partial\bar\mu^n}N=0$ for $n\ge 2$. Strictly speaking this
is correct
for a two dimensional system; for $d\ne 2$ we introduce errors of order
$\Delta\over\bar\mu$ or less.
\par Substituting the expansion, Eq. (40), into the expression for $Z_C$, Eq.
(35), we
obtain
\be Z^{{\i}}_C(N,T,\Phi)={\beta\over 2\pi i}\, e^{-\beta
F^{{\i}}(\bar\mu)}\int\limits_{\tilde C}\! d\mu e^{{\beta\over
2\Delta}(\mu-\bar\mu)}
\cdot \exp\left\{\beta\sum^\infty_{n=1}{1\over n!}{\partial^{n-1}\delta
N^{{\i}}\over\partial\bar\mu^{n-1}}(\mu-\bar\mu)^n\right\}\; .\ee 
We now define new variables
\beq \tau &\equiv& -i\sqrt{\beta\over\Delta}(\mu-\bar\mu)\nonumber \\
V^{{\i}}_n &\equiv& {1\over n!}{(i)^n\over (\beta\Delta)^{{n\over 2}-1}}
\Delta^{n-1}
{\partial^{n-1}\over\partial\bar\mu^{n-1}}\delta N^{{\i}}(\bar\mu)\;
,\eeq
in terms of which we may rewrite the partition function as
\be Z^{{\i}}_C(N,T,\Phi)=\sqrt{\Delta\beta\over 2\pi}\, e^{-\beta
F^{{\i}}(\bar\mu)}
\int\limits^\infty_{-\infty}\! {d\tau\over\sqrt{2\pi}} \, e^{-\tau^2\over 2}\,
\exp\left\{\sum^\infty_{n=1} V^{{\i}}_n\tau^n\right\}\; ,\ee
Equation (43) is the basis of our diagrammatic expansion. We first stress that all
diagrams
in this expansion are to be evaluated within the {\it \gc} ensemble, which is
the
attractive feature of our analysis. The vertex $V_n$ should be understood in
the
following way: $\delta N$ is a trace of a ring diagram with one scalar vertex
(i.e.
proportional to the first derivative of $\Omega$ with respect to $\bar\mu$).
Hence
$V_n$ corresponds to $n$ scalar vertices. (One may dress such vertices by
interaction
lines, impurity lines, etc. as depicted in Fig. 5). We now expand the
exponential in
Eq. (43). We shall comment on the small parameter of the expansion below. In
analogy
with Wick's theorem we are now able to define contraction with the variable
$\tau$. We
shall employ the following notation: for the second order term in $\tau$ we
write
\be \la \overbrace{\tau\tau}\ra\equiv \int\!{d\tau\over\sqrt{2\pi}}\,
e^{-\tau^2\over
2}\cdot \tau^2 =1\;.\ee 
This defines the combinational factor that comes with this term (=1),
corresponding to
the fact that there is only one way of contraction. For the fourth order term
in $\tau$
we write
\be \la \tau\tau\tau\tau\ra\equiv \int\!{d\tau\over\sqrt{2\pi}}\,
e^{-\tau^2\over
2}\tau^4 =3\; .\ee
The combinational factor associated with this term is 3, corresponding  to 3
possibilities of contraction. In these ``contractions" the role of the zeroth
order
propagation is played by $e^{-\tau^2\over 2}$. In our diagrammatic
representation we
shall use a zig-zag line for this propagator. We stress that the only role of
this
statistical $\tau$-propator is to account for combinatorial factors. It does
not carry
any momentum or energy. We also note that odd powers of $\tau$ vanish.
\par To evaluate \td\ derivatives we need to calculate {\it log\/} $Z_C$. To this end
we may
employ the linked cluster expression (including statistical lines!) whereby
only linked
diagrams are to be accounted for (cf. e.g. Ref. 29). We thus may write
\be F^{{\i}}(N,T,\Phi)=F^{{\i}}_{TD} (\bar\mu,T,\Phi) -{1\over\beta}\,log
\sqrt{\Delta\beta\over 2\pi} -\{all\, connected\,
diagrams\}^{{\i}}\;.\ee
Anticipating the discussion of the next section, partaining to differences
between
statistical mechanical and \td\ quantities, we use the notation
\be  F^{{\i}}_{TD}(\bar\mu,T,\Phi)\equiv\Omega^{{\i}}(\bar\mu, T,\Phi)+\bar\mu
N\, .\ee
The first term on the r.h.s. of Eq. (46) yields the sample specific \gc\
contribution.
The second term, due to Gaussian fluctuations around the expansion point (not
the
saddle point!) $\bar\mu$, is a temperature-dependent sample-independent
contribution.
All the other terms are, in principle, sample specific. The quantity
$\Omega^{{\i}}(\bar\mu,T,\Phi)$ may be evaluated employing the standard linked
cluster
expansion [30], which does not include statistical lines. The r.h.s. of Eq.
(46) yields
terms which are particular to the canonical ensemble.
\par At this point we can formulate the rules for constructing canonical
diagrams. We
consider a linked diagram consisting of $p$ electronic loops (with
$n_1,n_2,\dots n_p$
scalar vertices respectively). The total number of vertices must be even
$n_1+n_2+\ldots n_p=2K$. As all these vertices are connected by statistical
lines,
it follows that $K$ is the number of such lines, $K\ge p-1$.
\newpage\noindent
\beq
1.& &\hbox {With each statistical line associate a factor
$(-\Delta)$.}\nonumber\\
 2.& &\hbox{With each electronic loop which contains $n_i$
vertices associate a factor}\nonumber\\
&~&\hbox{ ${1\over n_i!}{\partial^{n_i-1}\over\partial
\bar\mu^{n_i-1}}\delta N(\bar\mu)$.}\nonumber \\
3.& & \hbox{Any subset of $m$ loops which consist of the same number
of}\nonumber \\
 &~&\hbox{vertices carries a factor ${1\over m!}$.}\nonumber \\
4.& & \hbox{Each diagram carries a factor $T^w, w=K-p+1$.}\nonumber \\
5.& & \hbox{With each diagram associate a combinatorial factor, representing
the}
\nonumber\\
 &~&\hbox{number of different ways of constructing it.}\eeq
We stress that the numerical factors obtained through the contractions
discussed above
correspond, in the present language, to the number of different ways of
connecting
scalar vertices by statistical lines. Hence, the only role of the statistical
lines is
to provide the right counting of diagrams. We also stress that the tempertaure
factor
$T^w$ associated with each diagram (see rule no. 4 above) does not exhaust the
temperature dependence of that particular diagram. The latter usually contains
further
important temperature dependence. We shall nevertheless use this power of $T$
to
classify the diagrams into different families, characterized by the index $w$.
Examples are depicted in Figs. 6 and 7. It is understood that disorder
averaging has to
be carried out subsequently. At zero temperature only the $w=0$ family
survives. It is
easy to see that these diagrams originate from the linear term in Eq. (40). An
expansion about the true saddle point, $\mu_0^{(i)}(T,\Phi)$ would not reproduce
these
diagrams. It turns out that these are the contributions which describe
differences
between the CE and the GCE at $T=0$, cf. Section 2. They are obtained when Eq.
(20) is
assumed to hold, and each member of the canonical ensemble is assigned an
effective,
sample specific chemical potential. The $w=0$ family represents contributions
due to
quenched disorder (as in Refs. [6-10,14]) but not due to differences between
statistical
mechanics and
\td s (see Section 3d). After some algebra one obtains (hereafter we
suppress the index $(i)$ and consider only ensemble averaged quantities)
\be \la\delta F_{w=0}\ra=-\sum^\infty_{n=1}\, {(-\Delta)^n\over
(n+1)!}{\partial^{n-1}\over\partial\bar\mu^{n-1}}\la\delta N^{n+1}\ra\, .\ee
Derivation of Eq. (59) with respect to an external parameter, $\Phi$, yields
Eq. (18)
[14]. The first $(n=1)$ term in the sum corresponds to the two loop diagram in
Fig. 6.
Upon averaging over (diffusive) disorder, it yields the Altshuler-Shklovskii
term [4],
which was employed in Ref. [6-10]. The $n\ge 2$ terms include complete
derivatives with
respect to $\bar\mu$, and are negligible upon averaging (being small in the
parameter
$\Delta/\bar\mu$). This provides us with {\it a posteriori} justification of
the
diagrammatic expansion: sample specific terms in this expansion are not
necessarily
small, but the ensemble average is well-behaved.
\par We next include the $w\ge 1$ families too. It can be shown that for the
regime
where we employ our expansion, $T>\Delta$, the leading term (in $\Delta/T$) of
each
family is represented by a two-loop diagram (cf. Figs. 6,7). Evaluation of
these
diagrams (neglecting full derivatives with respect to $\bar\mu$) results in
\be \la\delta F_{two-loop}\ra   ={\Delta\over 2}\sum^\infty_{w=0}
{(\Delta T)^w\over(w+1)!} \bigg\la\bigg({\partial^w\delta
N\over\partial\bar\mu^w}\bigg)^2\bigg\ra\; .\ee 
The $w=0$ term is the Altshuler-Shklovskii quenched disorder contribution
[4].\par
\bno
{\bf 3c~~~Relation to Spectral Correlations: Persistent Current and Heat
Capacity}
\par The summation in Eq. (50) is clearly related to correlations in the
spectra of
finite size disordered systems. We define the correlator $\cal K$ and its
Fourier
transform
$\tilde {\cal K}$:
\be {\cal K}(\epsilon-\epsilon')\equiv\Delta^2
\big(\la\nu(\epsilon)\nu(\epsilon')\ra\big)-
\la\nu(\epsilon)\ra\la(\epsilon')\ra\big)\equiv {1\over
2\pi}\int\limits^{+\infty}_{-\infty}\! dt\tilde {\cal K}(t)
e^{it(\epsilon-\epsilon')/\Delta}\; ,\ee
where $\nu(\epsilon)$ is the sample specific density of states and $t$ is the
dimensionless time (in units of $\hbar/\Delta$). We can write
\be \bigg\la\bigg({\partial^w\delta N\over\partial\bar\mu^w}\bigg)^2\bigg\ra =
{\partial^w\over\partial\bar\mu^w}{\partial^w\over\partial\bar\mu'^w}
\int\!\!\int\limits^{+\infty}_{-\infty}\!{d\epsilon d\epsilon'\over\Delta^2}\,
f(\epsilon-\bar\mu)
f(\epsilon'-\bar\mu'){\cal K}(\epsilon-\epsilon')\bigg|_{\mu'=\mu}\;
,\ee
where $f$ is the Fermi-Dirac function. Changing variables to $\xi\equiv
\epsilon-\epsilon'$, $\eta\equiv (\epsilon+\epsilon')/2$, performing the
integral
over $\eta$, and Fourier transforming with respect to $\xi$, we obtain
\be \la\delta F_{two-loop}\ra ={\pi T\over 2}\sum^\infty_{w=0} {1\over
(w+1)!}\bigg({\Delta\over T}\bigg)^w\int\limits^\infty_0\!
dt{t^{2w}\over\sinh^2\pi t}
\tilde {\cal K}\bigg(t{\Delta\over T}\bigg)\; .\ee  
Equations (46) and (53) form the basis for the analysis of the various corrections
to the
GCE averages, and depict the dependence of these corrections on spectral
correlations.
We study two examples comparing the $w=0$ (quenched disorder contribution) and
$w=1$.
We consider the case of diffusive disorder.
\par The average {\it persistent current} in the canonical ensemble, $\la
I\ra_{CE}$, is
obtained by deriving $\la F\ra$ with respect to the Aharonov-Bohm flux,
$\Phi$. The
flux dependent part of the time correlator for quasi-one-dimensiional rings is
given by
\be \tilde {\cal K}(t)={|t|\over\pi}\sqrt{1\over 4\pi g|t|}  \sum^\infty_{p=1}
e^{-{p^2\over 4g|t|}}\cos 4\pi p\Phi/\Phi_0\;,\ee
where $g=E_c/\Delta\gg 1$ is the dimenionless conductance ($E_c$ is the
Thouless
correlation energy) and $\Phi_0=hc/e$. Only even harmonics appear in Eq.
(54).
Expanding $\la I\ra_{CE}=\Sigma^\infty_{p=1} I_p\sin 4\pi p\Phi/\Phi_0$, we
find that
the quenched disorder contribution ($w=0$) is [6-10]
\be
I^{dis}_p={\Delta\over\Phi_0}\cases{{2\over\pi}& $\Delta<T<E_c/p^2 \; ,$ \cr
p^2 {T\over E_c}\, e^{-\sqrt{2\pi T\over E_c}p} & $T>E_c/p^2 \; .$\cr}
\ee
We shall denote the leading $w=1$ contribution with a superscript $SM-TD$. The
rationale behind that will become evident in the next section. We obtain
\be I^{SM-TD}_p ={\Delta\over \Phi_0}{1\over
g}\cases{15\sqrt{2\pi}{\zeta(5/2)\over
64\pi^3}p\sqrt{E_c\over T} &$\Delta<T<E_c/p^2\; ,$\cr
\noalign{\smallskip}
{1\over 16\pi} p^4{T\over E_c} e^{-\sqrt{{2\pi T\over E_c}}p}& $T>E_c/p^2\;
.$\cr}\ee
\par The average canonical {\it heat capacity} $(\la C\ra_{CE})$. It is given
by $\la
C\ra_{CE}=-\beta^2\partial^2(\beta\la F\ra)/\partial\beta^2$ and is written as
a sum of
contributions (cf. Eq. (46):
\be \la C\ra_{CE}=\la C\ra_{GCE}+\la\delta C^{Gauss}\ra +\la\delta C^{dis}\ra
+
\la\delta C^{SM-TD}\ra\; .\ee
The first term is the \gc\ contribution, which for a degenerate gas of
non-interacting
electrons is ${\pi^2\over 3}{T\over D}$. The term due to Gaussian fluctuations
yields
$-{1\over 2}$. This contribution can be reinterpreted when $T>\Delta$ as a
shift of the
\gc\ temperature $(T>\Delta)$ towards a lower temperature $T\rarrow T-{3\over
2\pi^2}\Delta$ forced by the canonical constraints. This is qualitatively
similar to the
shift found in Section 3a, in the context of the single level occupation
function. A
similar shift in the heat capacity was found in Ref. [20] (see also Refs.
[21,22]). The
next terms in Eq. (57) are evaluated employing Eq. (53). We are interested in
energies
larger than
$\Delta$, hence $t<1$. The time correlator has two interesting regimes (random
matrix
theory and Altshuler-Shklovskii [4] respectively):
\be \tilde K(t)={|t|\over b\pi}\cases{ 1&$g^{-1}<t<1$\cr
(4\pi g|t|)^{-d/2}&$\Delta\tau_{e\ell}<t<g^{-1}\; ,$\cr}\ee
where $b=1,2,4$ for the orthogonal, unitary and symplectic ensembles
respectively. Here
$\tau_{e\ell}$ is the elastic mean free time. Eq. (58) does not account for
the
crossover regimes. It leads to
\be\la\delta C^{dis}\ra ={1\over b\pi}{\Delta\over T}\cases{ {1\over 2\pi} &
$\Delta
<T< E_c$\cr
\gamma_d\bigg({T\over E_c}\bigg)^{d/2}  & $T>E_c$\cr}\ee
and in the leading order in $\Delta/T$ [31]
\be\la\delta C^{SM-TD}\ra={1\over b\pi}\bigg({\Delta\over
T}\bigg)^2\cases{{3\zeta(3)\over 4\pi} & $\Delta<T<E_c$\cr
\eta_d\bigg({T\over E_c}\bigg)^{d/2} & $T>E_c\; .$\cr}\ee
In this example the $SM-TD$ contribution is parametrically smaller than the
contribution due to quenched disorder.\par
\bno
{\bf 3d~~~Differences Between Statistical Mechanis and Thermodynamics}
\par Previous studies of differences between the GCE and the CE used the \td\
identity,
Eq. (20) as a starting point. This implies that for any specific realization,
a
canonical system can be replaced by an equivalent \gc\ system with an
effective
chemical potential which depends on disorder and on the value of the relevant
external
parameters $(\Phi)$. Physical observables, though, should be evaluated within
the
framework of \sm. While Eq. (20), considered as a \td\ relation, is exact --
by
construction, a \sm\ relation between $F$ and $\Omega$ contains important
corrections,
cf. Eq. (46). These corrections are expected to be suppressed in the ``\td\
limit" -- at high enough temperatures and/or large enough system's size.
\par These types of corrections (deviations from the thermodynamic relation,
Eq. (20))
may be quite important on the \msc\ scale. Thus, the expansion outlined above
includes
two qualitatively different classes of contributions: $(a)$ the $w=0$ family,
which
assumes the validity of Eq. (20), and accounts for differences between the CE
and the
GCE due to quenched disorder; $(b)$ the $w\ge 1$ families, which account for
deviations (within \sm) from the \td\ identities. (This is the rationale
behind the
notation used in Eqs. (56), (57) and (60)). The two-loop term of the $w=1$
family is
the leading contribution  (in $\Delta/T$) of this class.
\par Coming now to the examples discussed in the previous section, we note
that the
``$SM-TD$" contributions are smaller than those due to quenched disorder. It
is
interesting to note, though, that the former is quite robust as function
termperature:
contrary to what might be a naive expectation, contributions due to
differences between
\sm\ are {\it not} suppressed on the scale $T\sim\Delta$. In the case of the
persistent current the ``$SM-TD$" contribution (Eq. (56)) decreases
algebraically with
temperature up to $T\sim E_c\gg\Delta$, and is then suppressed exponentially.
In the
case of the heat capacity it is also suppressed algebraically (up  to
$T\sim\hbar/\tau_{e\ell}$. Our analysis
underlines the relation between the ``$SM-TD$" contributions and spectral
correlations.\par
\vskip .7in\noindent
{\bf Acknowledgements}
\par We have benefited from discussion with J. Hajdu, Y. Imry, A. Schmid and
A.D.
Stone. This research was supported by the German-Israel Foundation (GIF), the
U.S.-Israel Binational Science Foundation (BSF) and the Israel Academy of
Sciences.

\newpage
\centerline{\bf Figure Captions}
\begin{description}
\item [Fig. 1:]Four different thought experiments leading to different
definitions of
$\delta N$ (see text).
\item[Fig. 2:]Fluctuations in the number of electrons done to quenched
disorder. (a)
$\delta N^{(3)}$: At zero temperature members (1), (2) and (3) of the ensemble
(differing
in their respective shapes, impurity configurations, etc.) will have 3, 4 and
2
electrons, respectively. (b)~~$\delta N^{(4)}$: Fluctuations in $N$ due to
variations
of an external parameter, $\Phi$. The equilibrium values of $N$ at
$\Phi_1,\,\Phi_2$
are 3,2, respectively.
\item[Fig. 3:]Empty-occupied levels near the Fermi energy, (a) GCE; (b) CE.
Enhanced
probability for chemical potential to fall in a narrow gap.
\item[Fig. 4:]A two-level system under (a) canonical, (b) \gc\ constraints.
There are
no observable differences at $T=0$.
\item[Fig. 5:] A contribution to $V_5$. Solid lines denote electron
propagators;
wiggly lines -- electron-electron interactions; dashed lines -- impurity
scattering.
\item[Fig. 6:] The $w=0$ family of skeleton canonical diagrams with $K=1,2,3$.
Zigzag
lines are ``statistical lines"; full lines -- electron propagators; black dots
--
scalar vertices. The first diagram yields the Altshuler-Shklovskii result upon
averaging
over disorder.
\item[Fig. 7:] Skeleton diagrams of the $w=1$ family (up to $K=3$).
\end{description}


\begin{thebibliography}{31}
\bibitem[3]{}For the ergodic hypothesis to be valid one has to require that
the time
difference between subsequent measurements of the number of electrons exceeds
the
equilibrium time of the system.
\bibitem[11]{}Most of our discussion ignores spin degeneracy. Final results
will
include trivial spin degeneracy factors.
\bibitem[12]{}Note that there may be other sources that lead to broadening,
e.g.,
coupling to a heat bath reservoir, without allowing for particle exchange.
\bibitem[28]{}We note that $\mu_0$ is real. In fact, there is an infinite
number of
stationary points at $\mu_0\pm 2\pi i{n\over\beta}$, $n$ being an integer. We
choose
our contour, though, such that only the stationary point at $\mu_0$
contributes.
\end{thebibliography}
\end{document}